\documentclass[10pt,aps,twocolumn,nobibnotes,pra,showpacs]{revtex4-1}
  \usepackage{color}
  \usepackage{newlfont}
  \usepackage{graphicx}
  \usepackage{amssymb}
  \usepackage{amsmath}
  \usepackage{bm}
  \usepackage{verbatim}
  \usepackage[latin1]{inputenc}
  \usepackage{bbm}
  \usepackage{mathptmx}
  	
\newcommand{\ket}[1]{\mbox{$ | #1 \rangle $}}
\newcommand{\bra}[1]{\mbox{$ \langle #1 | $}}
\newcommand{\be}{\begin{eqnarray}}
\newcommand{\ee}{\end{eqnarray}}
\begin{document}

\title{Quantification of Einstein-Podolski-Rosen steering for two-qubit states}
\author{A. C. S. Costa}
\author{R. M. Angelo}
\affiliation{Department of Physics, Federal University of Paran\'a, P.O.Box 19044, 81531-980, Curitiba, PR, Brazil}

\begin{abstract}
In the last few years, several criteria to identify Einstein-Podolski-Rosen steering have been proposed and experimentally implemented. On the operational side, however, the evaluation of the steerability degree of a given state has shown to be a difficult task and only a few results are known. In this work, we propose a measure of steering that is based on the maximal violation of well established steering inequalities. Applying this approach to two-qubit states, we managed to derive simple closed formulas for steering in the two- and three-measurement scenarios. We also provide closed formulas for quantifiers of Bell nonlocality in the respective scenarios. Finally, we show that our measures of steering verify the entanglement-steering-nonlocality hierarchy and reproduce results reported so far. 
\end{abstract}


\maketitle

{\em Introduction}. The notion of {\em steering} was introduced by Schr\"odinger in 1935~\cite{schrodinger35,schrodinger36}, within the context of the Einstein-Podolski-Rosen (EPR) paradox, to name the ability of an observer to affect the state of a far remote system through local measurements. Specifically, if Alice and Bob share an entangled state, by performing measurements only in her part of the system, she can remotely steer Bob's state. This is not possible if the shared state is only classically correlated. This kind of quantum correlation is today known as EPR steering (or, simply steering).

Recently, steering was given an operational interpretation in a context in which Alice wants to persuade Bob, who does not trust her, that they share an entangled state~\cite{wiseman07,jones07}. In this scenario, these works established a hierarchy according to which steering lies in between entanglement~\cite{horodecki09} and Bell nonlocality \cite{brunner14}, this meaning that not every entangled state is steerable and not every steerable state is Bell nonlocal. The proof that entanglement, steering, and nonlocality are inequivalent when general measurements are considered was given only posteriorly~\cite{quintino15}.

Steerable states were shown to be advantageous for tasks involving randomness generation~\cite{law14}, subchannel discrimination~\cite{piani15}, quantum information processing~\cite{branciard11}, and one-sided device-independent processing in quantum key distribution~\cite{branciard12}, within which context a resource theory of steering has recently been formulated ~\cite{gallego15}. In addition, it has been proved that any set of incompatible measurements can be used for demonstrating EPR steering~\cite{quintino14,uola14} and that steering is fundamentally asymmetrical~\cite{bowles14}. On the experimental side, verifications of steerability have been reported for Bell-local entangled states~\cite{saunders10}, entangled Gaussian modes of light~\cite{handchen12}, and also in loophole-free experiments~\cite{wittmann12,smith12,bennet12}. 

Most of the aforementioned works diagnose steerability by testing necessary conditions, which are usually stated in terms of inequalities. The contributions of Reid~\cite{reid89} and Cavalcanti {\em et al}~\cite{cavalcanti09} define a seminal framework to this approach. Steering inequalities based on entropic uncertainty relations have also been proposed and experimentally tested~\cite{walborn11,schneeloch13,schneeloch13-2}. Further tools have been proposed to signalize steering, as for instance the all-versus-nothing proof without inequalities~\cite{chen13}, steering witnesses~\cite{kogias15-2}, Clauser-Horne-Shimony-Holt (CHSH)-like inequality~\cite{cavalcanti15,roy15}, and geometric Bell-like inequalities~\cite{zukowski15}.

Despite the vast knowledge accumulated so far with regard to witnessing steering, it is remarkable that, unlike nonlocality and entanglement, for which simple measures exist at least for some particular contexts~\cite{horodecki95,hill97,wootters98}, there is still scarce literature concerning the quantification of the steerability degree of a given quantum state. Developments along this line consist of the {\em steering weight}~\cite{skrzypczyk14}, whose evaluation demands the use of semidefinite programming (which is also the case for the computation of the {\em steering robustness}~\cite{piani15}), and a measure of steering for arbitrary bipartite Gaussian states of continuous variable systems~\cite{kogias15}. In particular, there is no closed formula even for the very important (and simple) case involving two-qubit states and a few measurements per site.

The aim of this Rapid Communication is to fill this gap. Confining ourselves to two- and three-measurement scenarios, we propose to quantify steering as the amount by which a given inequality is maximally violated. Focusing on some well-established steering inequalities, we derive measures that verify the entanglement-steering-nonlocality hierarchy and reproduce known results.

{\em Steering inequalities}. We start by collecting some results concerning the detection of steering. In a seminal paper~\cite{cavalcanti09}, Cavalcanti, Jones, Wiseman, and Reid (CJWR) developed an inequality to diagnose whether a bipartite state is steerable when Alice and Bob are both allowed to measure $n$ observables in their sites. This inequality is composed of a finite sum of bilinear expectation values:
\be
F_n^{\text{\tiny CJWR}}(\rho,\mu) =\frac{1}{\sqrt{n}} \Big|\sum_{i=1}^n\langle A_i\otimes B_i\rangle\Big| \leqslant 1,
\label{Iln}
\ee
where $A_i = \hat{u}_i\cdot\vec{\sigma}$,  $B_i = \hat{v}_i\cdot\vec{\sigma}$, $\vec{\sigma}=(\sigma_1,\sigma_2,\sigma_3)$ is a vector composed of the Pauli matrices, $\hat{u}_i\in\mathbb{R}^3$ are unit vectors, $\hat{v}_i\in\mathbb{R}^3$ are orthonormal vectors, $\mu=\{\hat{u}_1,\cdots,\hat{u}_n,\hat{v}_1,\cdots,\hat{v}_n\}$ is the set of measurement directions, $\langle A_i\otimes B_i\rangle=\text{Tr}(\rho A_i\otimes B_i)$, and $\rho\in\mathcal{H}_A\otimes\mathcal{H}_B$ is some bipartite quantum state. More recently, Cavalcanti {\em et al.}~\cite{cavalcanti15} considered a scenario in which Alice performs two dichotomic measurements while Bob performs two mutually unbiased qubit measurements. These authors then derived the following CHSH-like steering inequality:
\be
F_2^{\text{\tiny CHSH}}(\rho,\mu)=\tfrac{1}{2}\Big[\sqrt{f_+(\rho,\mu)} + \sqrt{f_-(\rho,\mu)}\Big]\leqslant 1,
\label{ICHSH}
\ee
where $f_{\pm}(\rho,\mu)=\langle (A_1\pm A_2)\otimes B_1\rangle^2 + \langle (A_1\pm A_2)\otimes B_2\rangle^2.$ It was recently shown~\cite{roy15} that the maximal value that the function $2 F_2^{\text{\tiny CHSH}}(\rho,\mu)$ can reach is $2\sqrt{2}$, which corresponds to Cirel'son's bound.

The aforementioned inequalities can be represented in the form $F_n(\rho,\mu)\leqslant 1$, where $F_n$ is some real-valued function, $\mu$ is a set of measurements, and $\rho$ is a bipartite state. Violations of these inequalities imply that $\rho$ is steerable for some $\mu$, but do not indicate how much steering this state possesses. 

{\em Steering measure}. Here we propose to quantify the degree of steerability of a given state by considering the amount by which a steering inequality is maximally violated. The rationale behind this strategy is identical to the one usually employed to quantify Bell nonlocality in bipartite states~\cite{brunner14,horodecki95}: a state that violates more an inequality, thus being more robust under noisy channels, is said to be more nonlocal (of course, this is not the only path for nonlocality quantification~\cite{comment1}). In this sense, our approach is also intuitively related to the notion of steering robustness~\cite{piani15}. We then propose the following measure of steering for a state $\rho$:
\be
S_n(\rho)&:=&\max\left\{0,\frac{F_n(\rho)-1}{F_n^{\text{\tiny max}}-1}\right\},
\label{Sgeneral}
\ee 
where 
\be 
F_n(\rho)=\max_{\mu}F_n(\rho,\mu)
\ee 
and $F_n^{\text{\tiny max}}=\max_{\rho}F_n(\rho)$. The normalization factor $F_n^{\text{\tiny max}}-1$ was introduced to render $S_n(\rho)\in[0,1]$. The inner maximization is taken over all measurement settings $\mu$, while the outer one selects maximal values that are greater than 1. Although Eq.~\eqref{Sgeneral} provides an intuitive operational measure of steering, this formulation still is mathematically involving due to the maximizations required. Nevertheless, we now show that analytical results can be obtained for {\em any} two-qubit state.

{\em Results}. In Ref.~\cite{luo08}, Luo showed that any two-qubit state can be reduced, by local unitary equivalence, to 
\be
\varsigma =\tfrac{1}{4}\Big(\mathbbm{1}\otimes\mathbbm{1}+\vec{a}\cdot\vec{\sigma}\otimes\mathbbm{1}+\mathbbm{1}\otimes\vec{b}\cdot\vec{\sigma}+\sum_{i=1}^3c_i\,\sigma_i\otimes \sigma_i\Big),
\label{rho}
\ee 
where $\mathbbm{1}$ is the $2\times 2$ identity matrix and $\{\vec{a},\vec{b},\vec{c}\}\in\mathbb{R}^3$ are vectors with norm less than unit. Because the state purity $P(\varsigma)\equiv \text{Tr}(\varsigma^2)=\tfrac{1}{4}\big(1+\vec{a}^2+\vec{b}^2+\vec{c}^2\big)$ is upper bounded by unit, we must have
\be
\vec{a}^2+\vec{b}^2+\vec{c}^2\leqslant 3. 
\label{abc<3}
\ee 
%

We now proceed to construct steering measures in accordance with the prescription \eqref{Sgeneral}. We start by the maximization of $F_n^{\text{\tiny CJWR}}(\varsigma,\mu)$ given in the inequality \eqref{Iln}. Using the state \eqref{rho}, we obtain the following expectation values:
\be
\langle\hat{u}_i\cdot\vec{\sigma}\otimes\hat{v}_i\cdot\vec{\sigma}\rangle &=&  \sum_{r=1}^3 u_{ir} c_r v_{ir} \equiv \bra{u_i}\mathcal{C}\ket{v_i} \equiv \mathcal{C}_i,
\label{meanvalues}
\ee
where $\mathcal{C}\equiv\sum_r c_r\ket{e_r}\bra{e_r}$ is a Hermitian operator with eigenvalues $c_r$, $\hat{u}_i=\ket{u_i} = \sum_r u_{ir}\ket{e_r}$, and $\hat{v}_i=\ket{v_i}=\sum_r v_{ir}\ket{e_r}$. For convenience, we have adopted bra-ket notation for the vectors $\{\hat{u}_i,\hat{v}_i\}$. Notice that $F_n^{\text{\tiny CJWR}}$ will not depend on either $\vec{a}$ or $\vec{b}$. Let us define 
\be
\ket{\alpha_i}\equiv\mathcal{C}\ket{u_i}
\label{|alphai>|ui>}
\ee 
such that $\mathcal{C}_i = \bra{\alpha_i}v_i\rangle$. This overlap is upper bounded by the norm $\alpha_i=\sqrt{\langle\alpha_i|\alpha_i\rangle}$ of the vector $\ket{\alpha_i}$. Aiming at maximizing $F_n^{\text{\tiny CJWR}}(\varsigma,\mu)=\left|\sum_i \mathcal{C}_i\right|/\sqrt{n}$ we then take all vectors $\ket{\alpha_i}$ and $\ket{v_i}$ to be parallel, that is,
\be
\ket{\alpha_i} = \alpha_i\ket{v_i}.
\label{|alphai>|vi>}
\ee
This can always be done, as we will discuss below. The orthonormality condition $\langle v_j|v_i\rangle=\delta_{ij}$ yields $\bra{\alpha_j}\alpha_i\rangle=\alpha_i^2\delta_{ij}$ and $\langle\alpha_j|v_i\rangle=\alpha_i\delta_{ij}$. Defining $\ket{\alpha}\equiv\sum_i\ket{\alpha_i}$ and $\ket{v}\equiv\sum_i \ket{v_i}$, we obtain that $\bra{\alpha}v\rangle = \sum_i \bra{\alpha_i}v_i\rangle$. This inner product is then an upper bound for $\sum_i\mathcal{C}_i$. Hence,
\be
F_n^{\text{\tiny CJWR}}(\varsigma,\mu)=\frac{1}{\sqrt{n}}\Big|\sum_i\langle\alpha_i|v_i\rangle\Big| \leqslant \frac{|\langle\alpha|v\rangle|}{\sqrt{n}}.
\ee
To find the maximum upper bound we impose that the vectors $\ket{\alpha}$ and $\ket{v}$ be parallel, that is,
\be
\ket{\alpha}/\alpha=\ket{v}/v, 
\label{|alpha>|v>}
\ee 
where $\alpha=\sqrt{\langle\alpha|\alpha\rangle}$ and $v=\sqrt{\langle v|v\rangle}=\sqrt{n}$ (by the  orthonormality of $\{\ket{v_i}\}$). Therewith, 
\be
F_n^{\text{\tiny CJWR}}(\varsigma,\mu)\leqslant\alpha,
\label{Fnl<alpha}
\ee
where
\be
\alpha^2=\sum_{i=1}^n\alpha_i^2=\sum_{i=1}^n \langle\alpha_i|\alpha_i\rangle = \sum_{i=1}^n \bra{u_i}\mathcal{C}^2\ket{u_i}.
\label{alpha2}
\ee
To determine $\alpha$, we first notice, by Eqs.~\eqref{|alphai>|ui>} and \eqref{|alphai>|vi>}, that
\be
\frac{\mathcal{C}\ket{u_i}}{\sqrt{\bra{u_i}\mathcal{C}^2\ket{u_i}}}=\ket{v_i}.
\label{|ui>|vi>}
\ee 
The orthonormality of $\{\ket{v_i}\}$ implies that the choice \eqref{|alphai>|vi>} will always be possible as long as $\bra{u_j}\mathcal{C}^2\ket{u_i}=\alpha_i^2\delta_{ij}$. This result requires $\ket{u_i}$ to be an eigenstate of $\mathcal{C}^2$ with eigenvalue $\alpha_i^2=c_i^2$. It follows, therefore, that $\alpha^2$ will be maximal if we choose vectors $\ket{u_i}$ that give the greatest eigenvalues of $\mathcal{C}$. With that, we finally obtain the maximal values
\be
F_2^{\text{\tiny CJWR}}(\varsigma) = \sqrt{c^2 - c^2_{\min}} \quad \text{and} \quad F_3^{\text{\tiny CJWR}}(\varsigma)=c,
\label{Fl}
\ee
where $c=\sqrt{\vec{c}^2}$ and $c_{\min} \equiv \min\{|c_1|,|c_2|,|c_3|\}$. 

It is worth noticing that these maximal values are tight. To show this, we start from $F_n^{\text{\tiny CJWR}}(\varsigma,\mu)=|\sum_i\mathcal{C}_i|/\sqrt{n}$ and use $|\sum_i\mathcal{C}_i|\leqslant \sum_i|\mathcal{C}_i|$, where $\mathcal{C}_i=\langle\alpha_i|v_i\rangle$. By the Cauchy-Schwarz inequality we have $|\langle\alpha_i|v_i\rangle|^2\leqslant \langle\alpha_i|\alpha_i\rangle=\alpha_i^2$. Now, because $(\alpha_i-\alpha_j)^2\geqslant 0$, which implies that $2\alpha_i\alpha_j\leqslant \alpha_i^2+\alpha_j^2$, one has that $(\sum_{i=1}^n\alpha_i)^2=\sum_i\alpha_i^2+2\sum_{i>j}\alpha_i\alpha_j\leqslant n\sum_{i=1}^n\alpha_i^2$, equality holding only if all $\alpha_i$ are equal. Thus far, we have then established the following tight inequalities:
\be
F_n^{\text{\tiny CJWR}}(\varsigma,\mu)=\tfrac{1}{\sqrt{n}}\Big|\sum_i\langle\alpha_i|v_i\rangle\Big|\leqslant \tfrac{1}{\sqrt{n}}\sum_i\alpha_i\leqslant \sqrt{\sum_i\alpha_i^2}.
\label{chain}
\ee 
Now, it is immediately seen that assumption \eqref{|alphai>|vi>} saturates the first inequality in the chain \eqref{chain}. By its turn, assumption \eqref{|alpha>|v>}, which means that $\sum_i\tfrac{\alpha_i}{\alpha}\ket{v_i}=\sum_i\tfrac{1}{\sqrt{n}}\ket{v_i}$, implies that $\alpha_i\sqrt{n}=\alpha$ for all $i$. It is easy to see that this relation saturates the last inequality in the chain \eqref{chain} and gives $\max_\mu F_n^{\text{\tiny CJWR}}(\varsigma,\mu)=\alpha$, in agreement with Eqs. \eqref{Fnl<alpha} and \eqref{Fl}.

We now proceed to the maximization of $F_2^{\text{\tiny CHSH}}(\varsigma,\mu)$ given in the inequality \eqref{ICHSH}. By Eq.~\eqref{meanvalues} one has $\langle A_i\otimes B_j\rangle=\bra{u_i}\mathcal{C}\ket{v_j}$, from which it follows that
\be
f_{\pm} = (\bra{u_1}\mathcal{C}\ket{v_1} \pm \bra{u_2}\mathcal{C}\ket{v_1})^2 + (\bra{u_1}\mathcal{C}\ket{v_2} \pm \bra{u_2}\mathcal{C}\ket{v_2})^2. \nonumber 
\ee
Since $\langle A_i\otimes B_j\rangle$ are real and $\sum_{i=1}^2\ket{v_i}\bra{v_i}=\mathbbm{1}$, one then shows that $f_{\pm} = \bra{u_1}\mathcal{C}^2\ket{u_1} + \bra{u_2}\mathcal{C}^2\ket{u_2} \pm 2\bra{u_1}\mathcal{C}^2\ket{u_2}$. Using $\ket{\alpha_i} \equiv \mathcal{C}\ket{u_i}$, we write $f_{\pm}=\alpha_1^2 + \alpha_2^2 \pm 2\alpha_1\alpha_2\cos\theta$, with $\theta$ the angular separation between $\ket{\alpha_1}$ and $\ket{\alpha_2}$. For fixed $\alpha_{1,2}$, maximization over $\theta$ gives $\sqrt{f}_++\sqrt{f}_-\leqslant 2(\alpha_1^2 + \alpha_2^2)^{1/2}$. The upper bound, which occurs for $\theta=\pi/2$, can always be reached by choosing $\{\ket{u_1},\ket{u_2}\}$ so as to verify $\langle\alpha_1|\alpha_2\rangle=0$. In this case, we can again define $\ket{\alpha}=\sum_{i=1}^2\ket{\alpha_i}$ and reduce the right-hand side of the above inequality to $\alpha=\sqrt{\langle\alpha|\alpha\rangle}$, which can be written as in Eq.~\eqref{alpha2}. We then obtain $F_2^{\text{\tiny CHSH}}(\varsigma,\mu)\leqslant \alpha$. Now it becomes obvious that we arrived precisely at the same optimization problem as defined by Eqs.~\eqref{Fnl<alpha} and \eqref{alpha2} with $n=2$. We then conclude that
\be
F_2^{\text{\tiny CHSH}}(\varsigma)=\sqrt{c^2-c_{\text{\tiny min}}^2}=F_2^{\text{\tiny CJWR}}(\varsigma).
\label{F2CHSH}
\ee 
We have numerically checked, over $10^7$ randomly generated settings $\mu$, that \eqref{Fl} and \eqref{F2CHSH} are indeed tight maximal values for their corresponding $F_n(\varsigma,\mu)$. To determine $F_n^{\text{\tiny max}}$, one may use the inequality \eqref{abc<3} with $\vec{a}^2=\vec{b}^2=0$ and the fact that $|c_i|\leqslant 1$ to obtain that $\max_{\varsigma}F_n^{\text{\tiny CJWR}}(\varsigma)=\sqrt{n}$. Thus, we come back to Eq.~\eqref{Sgeneral} to derive our final formula for the two-qubit steering:
\be\label{Sn}
&\displaystyle S_n(\varsigma)=\max\left\{0, \frac{\Lambda_n-1}{\sqrt{n}-1}\right\},& \\ \nonumber \\
&\Lambda_2=\sqrt{c^2-c_{\text{\tiny min}}^2}, \qquad \Lambda_3=c,& \nonumber 
\ee
which holds for any two-qubit state of form \eqref{rho} in settings involving $n=2,3$ measurements per site.
Notice that $S_n(\varsigma)$ does not depend on either $\vec{a}$ or $\vec{b}$. Interestingly, our results show that the proposal of quantifying steering via maximal violation unifies the notions of steering deriving from inequalities that are {\em a priori} very different, as \eqref{Iln} and \eqref{ICHSH}. 

{\em Entanglement, steering, and nonlocality}. We are now interested in assessing what hierarchy is implied by our measures of steering in relation to pertinent quantifiers of entanglement and nonlocality. To this end, we construct a measure of {\em CHSH nonlocality}, i.e., a quantifier for the maximal violation of the CHSH inequality, which refers to a scenario involving two measurements per site. Horodecki {\em et al} derived a necessary and sufficient condition for the violation of this inequality by generic mixed spin-1/2 states~\cite{horodecki95}. Defining a nonnegative quantity $M_{T_{\rho}}$ as a function of a matrix $T_{\rho}$ with elements $t_{ij}=\text{Tr}(\rho\sigma_i\otimes\sigma_j)$ and showing that $B_{\text{\tiny max}}=\max_{\nu}\langle B_{\text{\tiny CHSH}}\rangle_{\rho}=2\sqrt{M_{T_{\rho}}}$, where the maximum is taken over all measurement settings $\nu=\{\hat{x}_{1,2},\hat{y}_{1,2}\}$, they proved that the CHSH inequality $|\langle B_{\text{\tiny CHSH}}\rangle_{\rho}|=|\text{Tr}(\rho B_{\text{\tiny CHSH}})|\leqslant 2$, with a Bell operator $B_{\text{\tiny CHSH}}=\hat{x}_1\cdot\vec{\sigma}\otimes(\hat{y}_1+\hat{y}_2)\cdot\vec{\sigma}+\hat{x}_2\cdot\vec{\sigma}\otimes(\hat{y}_1-\hat{y}_2)\cdot\vec{\sigma}$, will be violated by $\rho$ if and only if $M_{T_{\rho}}>1$. Of particular importance here is the fact that, for a symmetric matrix $T_{\rho}$, $M_{T_{\rho}}=t_1^2+t_2^2$, where $|t_1|\geqslant |t_2|\geqslant |t_3|$ are the eigenvalues of $T_{\rho}$. By direct comparison of the state \eqref{rho} with the state (1) defined in Ref.~\cite{horodecki95}, we infer that $T_{\varsigma}\stackrel{.}{=}\{c_1,c_2,c_3\}$ (a diagonal matrix), which implies that $M_{T_{\varsigma}}=c^2-c_{\text{\tiny min}}^2$. Now, following the reasoning of the previous section and inspired by the results reported in Ref.~\cite{hu13}, we define our quantifier of CHSH nonlocality as the amount by which the aforementioned inequality is maximally violated, i.e., $N_2(\rho)\propto \max \left\{0,B_{\text{\tiny max}}-2\right\}$. Adopting a convenient normalization, we arrive at
\be 
N_2(\varsigma)=\max\left\{0,\frac{\sqrt{c^2-c^2_{\text{\tiny min}}}-1}{\sqrt{2}-1}\right\},
\label{N2}
\ee 
which is manifestly equal to $S_2(\varsigma)$. This shows that the notions of steering and Bell nonlocality here derived are indistinguishable in the two-measurement scenario. 

Let us compare $S_2(\varsigma)=N_2(\varsigma)$ with $S_3(\varsigma)$. From Eqs. \eqref{Sn} and \eqref{N2} it is easy to see that these correlations will be present if and only if $c^2>1+c^2_{\text{\tiny min}}$ and $c^2>1$, respectively. Hence, as the former inequality implies the latter, all CHSH-nonlocal states will necessarily be steerable (symbolically, $N_2\Leftrightarrow S_2\Rightarrow S_3$). However, not every state that is three-steerable (i.e., steerable in a three-measurement scenario) will be two-steerable and CHSH-nonlocal. For the comparison with entanglement, we invoke the {\em concurrence}, which for a two-qubit state $\rho$ reads $E(\rho)=\max\{0,\sqrt{\lambda_1}-\sqrt{\lambda_2}-\sqrt{\lambda_3}-\sqrt{\lambda_4}\}$, with $\lambda_1 \geqslant \lambda_2 \geqslant \lambda_3 \geqslant \lambda_4$ the eigenvalues of $\rho(\sigma_2 \otimes \sigma_2)\rho^*(\sigma_2 \otimes \sigma_2)$, and $\rho^*$ the complex conjugate of $\rho$ written in the computational basis~\cite{hill97}. In what follows, we will employ a very useful result, recently proved by Jafarpour and Sabour~\cite{sabour12}, according to which the concurrence of any two-qubit state is lower bounded by the concurrence of a related $X$ state. We can write an arbitrary two-qubit $X$ state $\rho_X$ as follows. Let $\rho_{jj}$ be the diagonal terms and $\rho_{jk}=|\rho_{jk}|e^{i\phi_{jk}}$ $(j<k)$ the antidiagonal ones, with $\rho_{kj}=\rho_{jk}^*$. It is a proven fact~\cite{yu07} that the concurrence of such a state can be written as
\be 
E(\rho_X)=2\max\left\{0,|\rho_{23}|-\sqrt{\rho_{11}\rho_{44}},|\rho_{14}|-\sqrt{\rho_{22}\rho_{33}}\right\}.\quad
\label{EX}
\ee 
Applying the {\em local} unitary transformation $U\equiv U_+\otimes U_-$, with $U_{\pm}=e^{-i\phi_{\pm}\sigma_3}$ and $\phi_{\pm}=\tfrac{1}{4}\left(\phi_{14}\pm\phi_{23}\right)$, one finds a five-parameter $X$ state of form
\be
U\rho_XU^{\dag}\stackrel{.}{=}\left( 
\begin{matrix}
\rho_{11} & 0 & 0 & |\rho_{14}| \\
0 & \rho_{22} & |\rho_{23}| & 0 \\
0 & |\rho_{23}| & \rho_{33} & 0 \\
|\rho_{14}| & 0 & 0 & \rho_{44}
\end{matrix}\right).
\label{rhoX}
\ee
The $X$ state \eqref{rhoX} can be parametrized as in Eq.~\eqref{rho} by suitably choosing matrix elements as $\rho_{jk}=\rho_{jk}(a_3,b_3,\vec{c})$. Therefore, up to a local unitary transformation that preserves every quantum correlation $Q$, any $X$ state can be written in the form \eqref{rho}, with $a_{1,2}=b_{1,2}=0$. Formally, $Q(\rho_X)=Q(\varsigma(a_3,b_3,\vec{c}))$. The Jafarpour-Sabour result~\cite{sabour12} then implies that
\be 
E(\varsigma(\vec{a},\vec{b},\vec{c}))\geqslant E(\varsigma(a_3,b_3,\vec{c})),
\label{E>EX}
\ee 
which indicates that the entanglement of any two-qubit state $\varsigma(\vec{a},\vec{b},\vec{c})$ is lower bounded by the concurrence of the associated $X$ state $\varsigma(a_3,b_3,\vec{c})$. Using formula \eqref{EX}, we can show that the entanglement of $\varsigma(a_3,b_3,\vec{c})$ can be written as $E(\varsigma(a_3,b_3,\vec{c})) = \tfrac{1}{2}\max\left\{0, \chi_+, \chi_-\right\}$, where $\chi_{\pm}\equiv|c_1\pm c_2|-[(1\pm c_3)^2-(a_3\pm b_3)^2]^{1/2}$. Since $(a_3\pm b_3)^2$ is non-negative, it follows that 
\be 
E(\varsigma(a_3,b_3,\vec{c}))\geqslant E(\varsigma(0,0,\vec{c})). 
\label{EX>EXc}
\ee 
Let $\lambda_1 \geqslant \lambda_2 \geqslant \lambda_3 \geqslant \lambda_4$ be the set of eigenvalues of $\varsigma(0,0,\vec{c})$, with $\lambda_1$ the greatest element. We can rewrite $\chi_{\pm}$ as
\be 
E(\varsigma(0,0,\vec{c}))=\max\{0,\mathfrak{e}\},\qquad \mathfrak{e}=2\lambda_1-1.
\label{Ee}
\ee 
By its turn, $S_3$ can be written in terms of the purity $P=\sum_i\lambda_i^2$ of $\varsigma(0,0,\vec{c})$ as
\be 
S_3(\varsigma(0,0,\vec{c}))=\max\{0,\mathfrak{s}\},\qquad \mathfrak{s}=\frac{\sqrt{4P-1}-1}{\sqrt{3}-1}.
\label{Ss}
\ee
To prove that nonseparability is a necessary condition for steerability, we use the fact that $(\sum_{i>1}\lambda_i)^2\geqslant \sum_{i>1}\lambda_i^2$. Since $P=\lambda_1^2+\sum_{i>1}\lambda_i^2$ and $\lambda_1+\sum_{i>1}\lambda_i=1$, we can rewrite this inequality as $P+2(\lambda_1-\lambda_1^2)\leqslant 1$. Given that $\mathfrak{e}=2\lambda_1-1$, it follows that $\mathfrak{e}\geqslant \sqrt{2P-1}$. By this inequality and Eq.~\eqref{Ss} we see that in order for $S_3>0$, which requires $P>\tfrac{1}{2}$, it is necessary that $E=\mathfrak{e}>0$, which proves the point. That entanglement can exist without steering can be shown by taking $c_i=-w$ with $w\in[0,1]$, which makes $\varsigma(0,0,\vec{c})$ reduce to a $2\times 2$ Werner state. Explicit calculation gives $\lambda_1=(1+3w)/4$ and $P=(1+3w^2)/4$. As a consequence, $\mathfrak{e}=(3w-1)/2$ and $\mathfrak{s}=(w\sqrt{3}-1)/(\sqrt{3}-1)$. We see that while steering only appears for $w>1/\sqrt{3}$, which is in full agreement with previously reported results~\cite{skrzypczyk14}, entanglement is already present for $w>1/3$. By inverting Eq.~\eqref{Ss} we obtain $P=P(\mathfrak{s})$ and, because $\mathfrak{e}\geqslant \sqrt{2P-1}$, we conclude that
\be 
E(\varsigma(0,0,\vec{c}))\geqslant S_3(\varsigma(0,0,\vec{c})).
\label{EXc>S3}
\ee  
By the relations \eqref{E>EX}, \eqref{EX>EXc}, and \eqref{EXc>S3}, we establish that
\be 
E(\varsigma)\geqslant S_3(\varsigma)
\label{E>S3}
\ee 
for any two-qubit state written in the form \eqref{rho}, which then means that $S_3\Rightarrow E$~\cite{comment2}. With that, we conclude that for the whole set of two-qubit states it holds that
\be 
N_2 \Leftrightarrow S_2 \Rightarrow S_3 \Rightarrow E,
\label{hierarchy}
\ee
implying a hierarchy according to which all CHSH-nonlocal states are steerable and all steerable states are entangled~\cite{wiseman07}. We numerically checked the hierarchy \eqref{hierarchy} over $10^7$ randomly generated states $\varsigma(0,0,\vec{c})$.

In light of the result $S_2(\varsigma)=N_2(\varsigma)$, the question arises whether $S_3$ would also be equivalent to a measure of Bell nonlocality involving {\em three} dichotomic measurements per site. We now prove that this equivalence does not exist. To this end, it is sufficient to focus on the Werner states $\rho_w=\varsigma(0,0,\vec{c})$ with $c_i=-w$ and $w\in[0,1]$. We construct a nonlocality measure, to be called $N_3$, that quantifies the maximal violation of the Bell-3322 inequality~\cite{collins04}, which can be stated as
\be 
I_{3322}&=&p(A_1B_1)+p(A_2B_1)+p(A_3B_1)+p(A_1B_2)\nonumber \\ &+& p(A_2B_2)-p(A_3B_2)+p(A_1B_3)-p(A_2B_3)\nonumber \\ &-&p(A_1)-p(B_2)-2p(B_1)\leqslant 0,
\label{I3322}
\ee 
where $p(A_iB_j)=\text{Tr}\left(M_i\otimes M_j\,\rho\right)$ and $p(A_k)=\text{Tr}(M_k\otimes\mathbbm{1}\,\rho)$ are probabilities associated with the von Neumann measurements $M_i=\tfrac{1}{2}(\mathbbm{1}+\hat{u}_i\cdot\vec{\sigma})$, for unit vectors $\hat{u}_i\in\mathbb{R}^3$.
Direct calculations yield $p(A_iB_j)=\tfrac{1}{4}(1-w\,\hat{u}_i\cdot\hat{v}_j)$ and $p(A_k)=\tfrac{1}{2}$.
From $(\vec{u}_i-\vec{v}_j)^2\geqslant 0$ we derive the inequality $\vec{u}_i\cdot\vec{v}_j\leqslant \tfrac{1}{2}\big(\vec{u}_i^2+\vec{v}_j^2\big)$, with which we can show that $I_{3322}\leqslant \frac{5w}{4}-1$.
Then, for $2\times 2$ Werner states, we have the following normalized measure of Bell-3322 nonlocality:
\be 
N_3(\rho_w)=\max\left\{0,5\,w-4\right\}.
\label{N3}
\ee 
According to Eq. \eqref{Sn}, $S_3(\rho_w)=\max\left\{0,\tfrac{w\sqrt{3}-1}{\sqrt{3}-1}\right\}$, which explicitly shows that $S_3$ and $N_3$ are inequivalent. At last, by direct inspection of the analytical results for $\rho_w$, we verify that $ N_3\Rightarrow N_2 \Leftrightarrow S_2 \Rightarrow S_3 \Rightarrow E$ for all $2\times 2$ Werner states. Clearly, the expected hierarchy is satisfied for $N_3$ as well. 

{\em Concluding remarks}. By looking at the maximal amount by which some steering inequalities are violated, we have derived closed formulas to quantify the steering of any two-qubit state of form \eqref{rho} in the two- and three-measurement scenarios. We also derived quantifiers of Bell nonlocality for each scenario. Besides correctly verifying the entanglement-steering-nonlocality hierarchy, our measures reproduce previously reported results. An open question is whether our strategy, as well as the robustness of steering \cite{piani15} and the steering weight~\cite{skrzypczyk14}, would lead to any sort of ``anomaly'' \cite{methot07} for bipartite states of higher dimension. In the affirmative case, it would be interesting to test further proposals, as for instance to compute steering by looking at the volume of violations in the parameter space \cite{fonseca15}.

{\em Acknowledgments}. This work was supported by CAPES and the National Institute for Science and Technology of Quantum Information (INCT-IQ/CNPq).



\begin{thebibliography}{70}

\expandafter\ifx\csname natexlab\endcsname\relax\def\natexlab#1{#1}\fi
\expandafter\ifx\csname bibnamefont\endcsname\relax
  \def\bibnamefont#1{#1}\fi
\expandafter\ifx\csname bibfnamefont\endcsname\relax
  \def\bibfnamefont#1{#1}\fi
\expandafter\ifx\csname citenamefont\endcsname\relax
  \def\citenamefont#1{#1}\fi
\expandafter\ifx\csname url\endcsname\relax
  \def\url#1{\texttt{#1}}\fi
\expandafter\ifx\csname urlprefix\endcsname\relax\def\urlprefix{URL }\fi
\providecommand{\bibinfo}[2]{#2}
\providecommand{\eprint}[2][]{\url{#2}}

\bibitem[{\citenamefont{Schr\"odinger}(1935)}]{schrodinger35}
\bibinfo{author}{\bibfnamefont{E.} \bibnamefont{Schr\"odinger}},
\bibinfo{author}{\bibfnamefont{Discussion of probability relations between separated systems}},
  \bibinfo{journal}{Math. Proc. Cambridge Philos. Soc.} \textbf{\bibinfo{volume}{31}},
  \bibinfo{pages}{555} (\bibinfo{year}{1935}).

\bibitem[{\citenamefont{Schr\"odinger}(1936)}]{schrodinger36}
\bibinfo{author}{\bibfnamefont{E.} \bibnamefont{Schr\"odinger}},
\bibinfo{author}{\bibfnamefont{Probability relations between separated systems}},
  \bibinfo{journal}{Math. Proc. Cambridge Philos. Soc.} \textbf{\bibinfo{volume}{32}},
  \bibinfo{pages}{446} (\bibinfo{year}{1936}).

\bibitem[{\citenamefont{Wiseman et al}(2007)}]{wiseman07} 
\bibinfo{author}{\bibfnamefont{H. M.} \bibnamefont{Wiseman}}, 
\bibinfo{author}{\bibfnamefont{S. J.} \bibnamefont{Jones}}, \bibnamefont{and}
\bibinfo{author}{\bibfnamefont{A. C.} \bibnamefont{Doherty}},
\bibinfo{author}{\bibfnamefont{Steering, Entanglement, Nonlocality, and the Einstein-Podolsky-Rosen Paradox}},
  \bibinfo{journal}{Phys. Rev. Lett.} \textbf{\bibinfo{volume}{98}},
  \bibinfo{pages}{140402} (\bibinfo{year}{2007}).

\bibitem[{\citenamefont{Jones et al}(2007)}]{jones07}
\bibinfo{author}{\bibfnamefont{S. J.} \bibnamefont{Jones}}, 
\bibinfo{author}{\bibfnamefont{H. M.} \bibnamefont{Wiseman}}, \bibnamefont{and}
\bibinfo{author}{\bibfnamefont{A. C.} \bibnamefont{Doherty}},
\bibinfo{author}{\bibfnamefont{Entanglement, Einstein-Podolsky-Rosen correlations, Bell nonlocality, and steering}},
  \bibinfo{journal}{Phys. Rev. A} \textbf{\bibinfo{volume}{76}},
  \bibinfo{pages}{052116} (\bibinfo{year}{2007}).
  
\bibitem[{\citenamefont{Horodecki et al}(2009)}]{horodecki09}
\bibinfo{author}{\bibfnamefont{R.} \bibnamefont{Horodecki}},
\bibinfo{author}{\bibfnamefont{P.} \bibnamefont{Horodecki}},
\bibinfo{author}{\bibfnamefont{M.} \bibnamefont{Horodecki}}, \bibnamefont{and}
\bibinfo{author}{\bibfnamefont{K.} \bibnamefont{Horodecki}}, 
\bibinfo{author}{\bibfnamefont{``Quantum entanglement,''}}
 \bibinfo{journal}{Rev. Mod. Phys.} \textbf{\bibinfo{volume}{81}},
 \bibinfo{pages}{865} (\bibinfo{year}{2009}).
  
\bibitem[{\citenamefont{Brunner et al}(2014)}]{brunner14}
\bibinfo{author}{\bibfnamefont{N.} \bibnamefont{Brunner}},
\bibinfo{author}{\bibfnamefont{D.} \bibnamefont{Cavalcanti}},
\bibinfo{author}{\bibfnamefont{S.} \bibnamefont{Pironio}},
\bibinfo{author}{\bibfnamefont{V.} \bibnamefont{Scarani}}, \bibnamefont{and}
\bibinfo{author}{\bibfnamefont{S.} \bibnamefont{Wehner}}, 
\bibinfo{author}{\bibfnamefont{``Bell nonlocality,''}}
 \bibinfo{journal}{Rev. Mod. Phys.} \textbf{\bibinfo{volume}{86}},
 \bibinfo{pages}{419} (\bibinfo{year}{2014}). 
  
\bibitem[{\citenamefont{Quintino et al}(2015)}]{quintino15}
\bibinfo{author}{\bibfnamefont{M. T.} \bibnamefont{Quintino}},
\bibinfo{author}{\bibfnamefont{T.} \bibnamefont{V\'ertesi}}, 
\bibinfo{author}{\bibfnamefont{D.} \bibnamefont{Cavalcanti}}, 
\bibinfo{author}{\bibfnamefont{R.} \bibnamefont{Augusiak}}, 
\bibinfo{author}{\bibfnamefont{M.} \bibnamefont{Demianowicz}},
\bibinfo{author}{\bibfnamefont{A.} \bibnamefont{Ac\'in}}, \bibnamefont{and}
\bibinfo{author}{\bibfnamefont{N.} \bibnamefont{Brunner}}, 
\bibinfo{author}{\bibfnamefont{Inequivalence of entanglement, steering, and Bell nonlocality for general measurements}},
  \bibinfo{journal}{Phys. Rev. A} \textbf{\bibinfo{volume}{92}},
  \bibinfo{pages}{032107} (\bibinfo{year}{2015}). 
  
\bibitem[{\citenamefont{Law et al}(2014)}]{law14}
\bibinfo{author}{\bibfnamefont{Y. Z.} \bibnamefont{Law}}, 
\bibinfo{author}{\bibfnamefont{L. P.} \bibnamefont{Thinh}},
\bibinfo{author}{\bibfnamefont{J.-D.} \bibnamefont{Bancal}}, \bibnamefont{and}
\bibinfo{author}{\bibfnamefont{V.} \bibnamefont{Scarani}}, 
\bibinfo{author}{\bibfnamefont{Quantum randomness extraction for various levels of characterization of the devices}},
  \bibinfo{journal}{J. Phys. A: Math. Theor.} \textbf{\bibinfo{volume}{47}},
  \bibinfo{pages}{424028} (\bibinfo{year}{2014}).

\bibitem[{\citenamefont{Piani et al}(2011)}]{piani15}
\bibinfo{author}{\bibfnamefont{M.} \bibnamefont{Piani}}, \bibnamefont{and}
\bibinfo{author}{\bibfnamefont{J.} \bibnamefont{Watrous}}, 
\bibinfo{author}{\bibfnamefont{Necessary and Sufficient Quantum Information Characterization of Einstein-Podolsky-Rosen Steering}},
  \bibinfo{journal}{Phys. Rev. Lett. } \textbf{\bibinfo{volume}{114}},
  \bibinfo{pages}{060404} (\bibinfo{year}{2015}). 
  
\bibitem[{\citenamefont{Branciard et al}(2011)}]{branciard11}
\bibinfo{author}{\bibfnamefont{C.} \bibnamefont{Branciard}} \bibnamefont{and}
\bibinfo{author}{\bibfnamefont{N.} \bibnamefont{Gisin}}, 
\bibinfo{author}{\bibfnamefont{Quantifying the Nonlocality of Greenberger-Horne-Zeilinger Quantum Correlations by a Bounded Communication Simulation Protocol}},
  \bibinfo{journal}{Phys. Rev. Lett. } \textbf{\bibinfo{volume}{107}},
  \bibinfo{pages}{020401} (\bibinfo{year}{2011}).  
  
\bibitem[{\citenamefont{Branciard et al}(2012)}]{branciard12}
\bibinfo{author}{\bibfnamefont{C.} \bibnamefont{Branciard}}, 
\bibinfo{author}{\bibfnamefont{E. G.} \bibnamefont{Cavalcanti}}, 
\bibinfo{author}{\bibfnamefont{S. P.} \bibnamefont{Walborn}},  
\bibinfo{author}{\bibfnamefont{V.} \bibnamefont{Scarani}}, \bibnamefont{and}
\bibinfo{author}{\bibfnamefont{H. M.} \bibnamefont{Wiseman}}, 
\bibinfo{author}{\bibfnamefont{One-sided device-independent quantum key distribution: Security, feasibility, and the connection with steering}},
  \bibinfo{journal}{Phys. Rev. A} \textbf{\bibinfo{volume}{85}},
  \bibinfo{pages}{010301(R)} (\bibinfo{year}{2012}).
  
\bibitem[{\citenamefont{Gallego et al}(2015)}]{gallego15}
\bibinfo{author}{\bibfnamefont{R.} \bibnamefont{Gallego}} \bibnamefont{and}
\bibinfo{author}{\bibfnamefont{L.} \bibnamefont{Aolita}}, 
\bibinfo{author}{\bibfnamefont{Resource Theory of Steering}},
  \bibinfo{journal}{Phys. Rev. X} \textbf{\bibinfo{volume}{5}},
  \bibinfo{pages}{041008} (\bibinfo{year}{2015}). 

\bibitem[{\citenamefont{Quintino et al}(2014)}]{quintino14}
\bibinfo{author}{\bibfnamefont{M. T.} \bibnamefont{Quintino}},
\bibinfo{author}{\bibfnamefont{T.} \bibnamefont{V\'ertesi}}, \bibnamefont{and}
\bibinfo{author}{\bibfnamefont{N.} \bibnamefont{Brunner}}, 
\bibinfo{author}{\bibfnamefont{Joint Measurability, Einstein-Podolsky-Rosen Steering, and Bell Nonlocality}},
  \bibinfo{journal}{Phys. Rev. Lett.} \textbf{\bibinfo{volume}{113}},
  \bibinfo{pages}{160402} (\bibinfo{year}{2014}).

\bibitem[{\citenamefont{Uola et al}(2014)}]{uola14}
\bibinfo{author}{\bibfnamefont{R.} \bibnamefont{Uola}},
\bibinfo{author}{\bibfnamefont{T.} \bibnamefont{Moroder}}, \bibnamefont{and}
\bibinfo{author}{\bibfnamefont{O.} \bibnamefont{G\"uhne}}, 
\bibinfo{author}{\bibfnamefont{Joint Measurability of Generalized Measurements Implies Classicality}},
  \bibinfo{journal}{Phys. Rev. Lett.} \textbf{\bibinfo{volume}{113}},
  \bibinfo{pages}{160403} (\bibinfo{year}{2014}).
  
\bibitem[{\citenamefont{Bowles et al}(2014)}]{bowles14}
\bibinfo{author}{\bibfnamefont{J.} \bibnamefont{Bowles}}, 
\bibinfo{author}{\bibfnamefont{T.} \bibnamefont{V\'ertesi}},
\bibinfo{author}{\bibfnamefont{M. T.} \bibnamefont{Quintino}}, \bibnamefont{and}
\bibinfo{author}{\bibfnamefont{N.} \bibnamefont{Brunner}}, 
\bibinfo{author}{\bibfnamefont{One-Way Einstein-Podolsky-Rosen Steering}},
  \bibinfo{journal}{Phys. Rev. Lett.} \textbf{\bibinfo{volume}{112}},
  \bibinfo{pages}{200402} (\bibinfo{year}{2014}).
  
\bibitem[{\citenamefont{Saunders et al}(2010)}]{saunders10}
\bibinfo{author}{\bibfnamefont{D. J.} \bibnamefont{Saunders}}, 
\bibinfo{author}{\bibfnamefont{S. J.} \bibnamefont{Jones}}, 
\bibinfo{author}{\bibfnamefont{H. M.} \bibnamefont{Wiseman}}, \bibnamefont{and}
\bibinfo{author}{\bibfnamefont{G. J.} \bibnamefont{Pryde}}, 
\bibinfo{author}{\bibfnamefont{Experimental EPR-steering using Bell-local states}},
  \bibinfo{journal}{Nature Phys.} \textbf{\bibinfo{volume}{6}},
  \bibinfo{pages}{845} (\bibinfo{year}{2010}).  
  
\bibitem[{\citenamefont{Handchen et al}(2012)}]{handchen12}
\bibinfo{author}{\bibfnamefont{V.} \bibnamefont{Handchen}}, 
\bibinfo{author}{\bibfnamefont{T.} \bibnamefont{Eberle}}, 
\bibinfo{author}{\bibfnamefont{S.} \bibnamefont{Steinlechner}}, 
\bibinfo{author}{\bibfnamefont{A.} \bibnamefont{Samblowski}}, 
\bibinfo{author}{\bibfnamefont{T.} \bibnamefont{Franz}}, 
\bibinfo{author}{\bibfnamefont{R. F.} \bibnamefont{Werner}}, \bibnamefont{and}
\bibinfo{author}{\bibfnamefont{R.} \bibnamefont{Schnabel}}, 
\bibinfo{author}{\bibfnamefont{Observation of one-way Einstein-Podolsky-Rosen steering}},
  \bibinfo{journal}{Nat. Photonics} \textbf{\bibinfo{volume}{6}},
  \bibinfo{pages}{596} (\bibinfo{year}{2012}).

\bibitem[{\citenamefont{Wittmann et al}(2012)}]{wittmann12}
\bibinfo{author}{\bibfnamefont{B.} \bibnamefont{Wittmann}}, 
\bibinfo{author}{\bibfnamefont{S.} \bibnamefont{Ramelow}}, 
\bibinfo{author}{\bibfnamefont{F.} \bibnamefont{Steinlechner}},  
\bibinfo{author}{\bibfnamefont{N. K.} \bibnamefont{Langford}}, 
\bibinfo{author}{\bibfnamefont{N.} \bibnamefont{Brunner}},
\bibinfo{author}{\bibfnamefont{H. M.} \bibnamefont{Wiseman}}, 
\bibinfo{author}{\bibfnamefont{R.} \bibnamefont{Ursin}}, \bibnamefont{and}
\bibinfo{author}{\bibfnamefont{A.} \bibnamefont{Zeilinger}}, 
\bibinfo{author}{\bibfnamefont{Loophole-free Einstein-Podolsky-Rosen experiment via quantum steering}},
  \bibinfo{journal}{New J. Phys.} \textbf{\bibinfo{volume}{14}},
  \bibinfo{pages}{053030} (\bibinfo{year}{2012}).

\bibitem[{\citenamefont{Smith et al}(2012)}]{smith12}
\bibinfo{author}{\bibfnamefont{D. H.} \bibnamefont{Smith}}, 
\bibinfo{author}{\bibfnamefont{G.} \bibnamefont{Gillett}}, 
\bibinfo{author}{\bibfnamefont{M. P.} \bibnamefont{de Almeida}},  
\bibinfo{author}{\bibfnamefont{C.} \bibnamefont{Branciard}}, 
\bibinfo{author}{\bibfnamefont{A.} \bibnamefont{Fedrizzi}},
\bibinfo{author}{\bibfnamefont{T. J.} \bibnamefont{Weinhold}}, 
\bibinfo{author}{\bibfnamefont{A.} \bibnamefont{Lita}},
\bibinfo{author}{\bibfnamefont{B.} \bibnamefont{Calkins}},
\bibinfo{author}{\bibfnamefont{T.} \bibnamefont{Gerrits}}, 
\bibinfo{author}{\bibfnamefont{H. M.} \bibnamefont{Wiseman}},
\bibinfo{author}{\bibfnamefont{S. W.} \bibnamefont{Nam}}, \bibnamefont{and}
\bibinfo{author}{\bibfnamefont{A. G.} \bibnamefont{White}}, 
\bibinfo{author}{\bibfnamefont{Conclusive quantum steering with superconducting transition-edge sensors}},
  \bibinfo{journal}{Nat. Commun.} \textbf{\bibinfo{volume}{3}},
  \bibinfo{pages}{625} (\bibinfo{year}{2012}).

\bibitem[{\citenamefont{Bennet et al}(2012)}]{bennet12}
\bibinfo{author}{\bibfnamefont{A. J.} \bibnamefont{Bennet}}, 
\bibinfo{author}{\bibfnamefont{D. A.} \bibnamefont{Evans}}, 
\bibinfo{author}{\bibfnamefont{D. J.} \bibnamefont{Saunders}},  
\bibinfo{author}{\bibfnamefont{C.} \bibnamefont{Branciard}}, 
\bibinfo{author}{\bibfnamefont{E. G.} \bibnamefont{Cavalcanti}}, 
\bibinfo{author}{\bibfnamefont{H. M.} \bibnamefont{Wiseman}}, \bibnamefont{and}
\bibinfo{author}{\bibfnamefont{G. J.} \bibnamefont{Pryde}}, 
\bibinfo{author}{\bibfnamefont{Arbitrarily Loss-Tolerant Einstein-Podolsky-Rosen Steering Allowing a Demonstration Over 1 km of Optical Fiber with No Detection Loophole}},
  \bibinfo{journal}{Phys. Rev. X} \textbf{\bibinfo{volume}{2}},
  \bibinfo{pages}{031003} (\bibinfo{year}{2012}).

\bibitem[{\citenamefont{Reid}(1989)}]{reid89}
\bibinfo{author}{\bibfnamefont{M. D.} \bibnamefont{Reid}},  
\bibinfo{author}{\bibfnamefont{Demonstration of the Einstein-Podolsky-Rosen paradox using nondegenerate parametric amplification}},
  \bibinfo{journal}{Phys. Rev. A} \textbf{\bibinfo{volume}{40}},
  \bibinfo{pages}{913} (\bibinfo{year}{1989}).

\bibitem[{\citenamefont{Cavalcanti et al}(2009)}]{cavalcanti09}
\bibinfo{author}{\bibfnamefont{E. G.} \bibnamefont{Cavalcanti}}, 
\bibinfo{author}{\bibfnamefont{S. J.} \bibnamefont{Jones}}, 
\bibinfo{author}{\bibfnamefont{H. M.} \bibnamefont{Wiseman}}, \bibnamefont{and}
\bibinfo{author}{\bibfnamefont{M. D.} \bibnamefont{Reid}}, 
\bibinfo{author}{\bibfnamefont{Experimental criteria for steering and the Einstein-Podolsky-Rosen paradox}},
  \bibinfo{journal}{Phys. Rev. A} \textbf{\bibinfo{volume}{80}},
  \bibinfo{pages}{032112} (\bibinfo{year}{2009}).
  
\bibitem[{\citenamefont{Walborn et al}(2011)}]{walborn11}
\bibinfo{author}{\bibfnamefont{S. P.} \bibnamefont{Walborn}}, 
\bibinfo{author}{\bibfnamefont{A.} \bibnamefont{Salles}}, 
\bibinfo{author}{\bibfnamefont{R. M.} \bibnamefont{Gomes}},  
\bibinfo{author}{\bibfnamefont{F.} \bibnamefont{Toscano}}, \bibnamefont{and}
\bibinfo{author}{\bibfnamefont{P. H.} \bibnamefont{Souto Ribeiro}}, 
\bibinfo{author}{\bibfnamefont{Revealing Hidden Einstein-Podolsky-Rosen Nonlocality}},
  \bibinfo{journal}{Phys. Rev. Lett.} \textbf{\bibinfo{volume}{106}},
  \bibinfo{pages}{130402} (\bibinfo{year}{2011}).  
 
\bibitem[{\citenamefont{Schneeloch et al}(2013)}]{schneeloch13}
\bibinfo{author}{\bibfnamefont{J.} \bibnamefont{Schneeloch}}, 
\bibinfo{author}{\bibfnamefont{C. J.} \bibnamefont{Broadbent}}, 
\bibinfo{author}{\bibfnamefont{S. P.} \bibnamefont{Walborn}},  
\bibinfo{author}{\bibfnamefont{E. G.} \bibnamefont{Cavalcanti}}, \bibnamefont{and}
\bibinfo{author}{\bibfnamefont{J. C.} \bibnamefont{Howell}}, 
\bibinfo{author}{\bibfnamefont{Einstein-Podolsky-Rosen steering inequalities from entropic uncertainty relations}},
  \bibinfo{journal}{Phys. Rev. A} \textbf{\bibinfo{volume}{87}},
  \bibinfo{pages}{062103} (\bibinfo{year}{2013}).
  
\bibitem[{\citenamefont{Schneeloch et al}(2013)}]{schneeloch13-2}
\bibinfo{author}{\bibfnamefont{J.} \bibnamefont{Schneeloch}}, 
\bibinfo{author}{\bibfnamefont{P.} \bibnamefont{Ben Dixon}}, 
\bibinfo{author}{\bibfnamefont{G. A.} \bibnamefont{Howland}},  
\bibinfo{author}{\bibfnamefont{C. J.} \bibnamefont{Broadbent}}, \bibnamefont{and}
\bibinfo{author}{\bibfnamefont{J. C.} \bibnamefont{Howell}}, 
\bibinfo{author}{\bibfnamefont{Violation of Continuous-Variable Einstein-Podolsky-Rosen Steering with Discrete Measurements}},
  \bibinfo{journal}{Phys. Rev. Lett.} \textbf{\bibinfo{volume}{110}},
  \bibinfo{pages}{130407} (\bibinfo{year}{2013}).
  
\bibitem[{\citenamefont{Chen et al}(2013)}]{chen13} 
\bibinfo{author}{\bibfnamefont{J.-L.} \bibnamefont{Chen}}, 
\bibinfo{author}{\bibfnamefont{X.-J.} \bibnamefont{Ye}}, 
\bibinfo{author}{\bibfnamefont{C.} \bibnamefont{Wu}}, 
\bibinfo{author}{\bibfnamefont{H.-Y.} \bibnamefont{Su}}, 
\bibinfo{author}{\bibfnamefont{A.} \bibnamefont{Cabello}},  
\bibinfo{author}{\bibfnamefont{L. C.} \bibnamefont{Kwek}},\bibnamefont{and}
\bibinfo{author}{\bibfnamefont{C. H.} \bibnamefont{Oh}},
\bibinfo{author}{\bibfnamefont{All-versus-nothing proof of Einstein-Podolsky-Rosen steering}},
  \bibinfo{journal}{Sci. Rep.} \textbf{\bibinfo{volume}{3}},
  \bibinfo{pages}{2143} (\bibinfo{year}{2013}).

\bibitem[{\citenamefont{Kogias et al}(2015)}]{kogias15-2}
\bibinfo{author}{\bibfnamefont{I.} \bibnamefont{Kogias}},
\bibinfo{author}{\bibfnamefont{P.} \bibnamefont{Skrzypczyk}},
\bibinfo{author}{\bibfnamefont{D.} \bibnamefont{Cavalcanti}},
\bibinfo{author}{\bibfnamefont{A.} \bibnamefont{Ac\'in}}, \bibnamefont{and}
\bibinfo{author}{\bibfnamefont{G.} \bibnamefont{Adesso}}, 
\bibinfo{author}{\bibfnamefont{Hierarchy of Steering Criteria Based on Moments for All Bipartite Quantum Systems}},
  \bibinfo{journal}{Phys. Rev. Lett.} \textbf{\bibinfo{volume}{115}},
  \bibinfo{pages}{210401} (\bibinfo{year}{2015}).  
  
\bibitem[{\citenamefont{Cavalcanti et al}(2014)}]{cavalcanti15}
\bibinfo{author}{\bibfnamefont{E. G.} \bibnamefont{Cavalcanti}},
\bibinfo{author}{\bibfnamefont{C. J.} \bibnamefont{Foster}},
\bibinfo{author}{\bibfnamefont{M.} \bibnamefont{Fuwa}}, \bibnamefont{and}
\bibinfo{author}{\bibfnamefont{H. M.} \bibnamefont{Wiseman}}, 
\bibinfo{author}{\bibfnamefont{Analog of the Clauser-Horne-Shimony-Holt inequality for steering}},
  \bibinfo{journal}{J. Opt. Soc. Am. B} \textbf{\bibinfo{volume}{32}},
  \bibinfo{pages}{A74} (\bibinfo{year}{2015}).

\bibitem[{\citenamefont{Roy et al}(2015)}]{roy15}
\bibinfo{author}{\bibfnamefont{A.} \bibnamefont{Roy}},
\bibinfo{author}{\bibfnamefont{S. S.} \bibnamefont{Bhattacharya}},
\bibinfo{author}{\bibfnamefont{A.} \bibnamefont{Mukherjee}}, \bibnamefont{and}
\bibinfo{author}{\bibfnamefont{M.} \bibnamefont{Banik}}, 
\bibinfo{author}{\bibfnamefont{Optimal quantum violation of Clauser-Horne-Shimony-Holt like steering inequality}},
  \bibinfo{journal}{J. Phys. A: Math. Theor.} \textbf{\bibinfo{volume}{48}},
  \bibinfo{pages}{415302} (\bibinfo{year}{2015}).

\bibitem[{\citenamefont{Zukowski et al}(2015)}]{zukowski15}
\bibinfo{author}{\bibfnamefont{M.} \bibnamefont{\.Zukowski}},
\bibinfo{author}{\bibfnamefont{A.} \bibnamefont{Dutta}}, \bibnamefont{and}
\bibinfo{author}{\bibfnamefont{Z.} \bibnamefont{Yin}}, 
\bibinfo{author}{\bibfnamefont{Geometric Bell-like inequalities for steering}},
  \bibinfo{journal}{Phys. Rev. A} \textbf{\bibinfo{volume}{91}},
  \bibinfo{pages}{032107} (\bibinfo{year}{2015}).  
  
\bibitem[{\citenamefont{Horodecki et al}(1995)}]{horodecki95}
\bibinfo{author}{\bibfnamefont{R.} \bibnamefont{Horodecki}},
\bibinfo{author}{\bibfnamefont{P.} \bibnamefont{Horodecki}}, \bibnamefont{and}
\bibinfo{author}{\bibfnamefont{M.} \bibnamefont{Horodecki}}, 
\bibinfo{author}{\bibfnamefont{Violating Bell inequality by mixed spin-1/2 states: Necessary and sufficient condition}},
 \bibinfo{journal}{Phys. Lett. A} \textbf{\bibinfo{volume}{200}},
 \bibinfo{pages}{340} (\bibinfo{year}{1995}).

\bibitem[{\citenamefont{Hill et al}(97)}]{hill97}
\bibinfo{author}{\bibfnamefont{S.} \bibnamefont{Hill}} \bibnamefont{and}
\bibinfo{author}{\bibfnamefont{M. K.} \bibnamefont{Wootters}},
\bibinfo{author}{\bibfnamefont{Entanglement of a Pair of Quantum Bits}},
 \bibinfo{journal}{Phys. Rev. Lett.} \textbf{\bibinfo{volume}{78}},
 \bibinfo{pages}{5022} (\bibinfo{year}{1997}).
 
\bibitem[{\citenamefont{Wootters}(1998)}]{wootters98}
\bibinfo{author}{\bibfnamefont{W. K.} \bibnamefont{Wootters}},
\bibinfo{author}{\bibfnamefont{Entanglement of Formation of an Arbitrary State of Two Qubits}},
 \bibinfo{journal}{Phys. Rev. Lett.} \textbf{\bibinfo{volume}{80}},
 \bibinfo{pages}{2245} (\bibinfo{year}{1998}).
  
\bibitem[{\citenamefont{Skrzypczyk et al}(2014)}]{skrzypczyk14}
\bibinfo{author}{\bibfnamefont{P.} \bibnamefont{Skrzypczyk}}, 
\bibinfo{author}{\bibfnamefont{M.} \bibnamefont{Navascu\'es}}, \bibnamefont{and}
\bibinfo{author}{\bibfnamefont{D.} \bibnamefont{Cavalcanti}}, 
\bibinfo{author}{\bibfnamefont{Quantifying Einstein-Podolsky-Rosen Steering}},
  \bibinfo{journal}{Phys. Rev. Lett.} \textbf{\bibinfo{volume}{112}},
  \bibinfo{pages}{180404} (\bibinfo{year}{2014}).
  
\bibitem[{\citenamefont{Kogias et al}(2015)}]{kogias15}
\bibinfo{author}{\bibfnamefont{I.} \bibnamefont{Kogias}},
\bibinfo{author}{\bibfnamefont{A. R.} \bibnamefont{Lee}},
\bibinfo{author}{\bibfnamefont{S.} \bibnamefont{Ragy}}, \bibnamefont{and}
\bibinfo{author}{\bibfnamefont{G.} \bibnamefont{Adesso}}, 
\bibinfo{author}{\bibfnamefont{Quantification of Gaussian Quantum Steering}},
  \bibinfo{journal}{Phys. Rev. Lett.} \textbf{\bibinfo{volume}{114}},
  \bibinfo{pages}{060403} (\bibinfo{year}{2015}). 
	
\bibitem{comment1} One can conceive, e.g., a scenario involving inefficient detectors, in which case a state that remains nonlocal for the largest detection inefficiency can be claimed to be more nonlocal. Interestingly, in such a framework, the singlet is no longer the most nonlocal state~\cite{eberhard93}. Alternatively, one can consider the {\em volume of violations} as the figure of merit for the degree of nonlocality~\cite{fonseca15}.

\bibitem[{\citenamefont{Eberhard}(93)}]{eberhard93}
\bibinfo{author}{\bibfnamefont{P. H.} \bibnamefont{Eberhard}},
\bibinfo{author}{\bibfnamefont{Background level and counter efficiencies required for a loophole-free Einstein-Podolsky-Rosen experiment}},
 \bibinfo{journal}{Phys. Rev. A} \textbf{\bibinfo{volume}{47}},
 \bibinfo{pages}{R747} (\bibinfo{year}{1993}).

\bibitem[{\citenamefont{Fonseca et al}(2015)}]{fonseca15}
\bibinfo{author}{\bibfnamefont{E. A.} \bibnamefont{Fonseca}} \bibnamefont{and}
\bibinfo{author}{\bibfnamefont{F.} \bibnamefont{Parisio}},
\bibinfo{author}{\bibfnamefont{Measure of nonlocality which is maximal for maximally entangled qutrits}},
 \bibinfo{journal}{Phys. Rev. A} \textbf{\bibinfo{volume}{92}},
 \bibinfo{pages}{030101(R)} (\bibinfo{year}{2015}).
  
\bibitem[{\citenamefont{Luo}(2008)}]{luo08}
\bibinfo{author}{\bibfnamefont{S.} \bibnamefont{Luo}},
\bibinfo{author}{\bibfnamefont{Quantum discord for two-qubit systems}},
 \bibinfo{journal}{Phys. Rev. A} \textbf{\bibinfo{volume}{77}},
 \bibinfo{pages}{042303} (\bibinfo{year}{2008}).
 
\bibitem[{\citenamefont{Hu}(2013)}]{hu13}
\bibinfo{author}{\bibfnamefont{M.-L.} \bibnamefont{Hu}},
\bibinfo{author}{\bibfnamefont{Relations between entanglement, Bell-inequality violation and teleportation fidelity for the two-qubit X states}},
 \bibinfo{journal}{Quantum Info. Process} \textbf{\bibinfo{volume}{12}},
 \bibinfo{pages}{229} (\bibinfo{year}{2013}).

\bibitem[{\citenamefont{Sabour et al}(2012)}]{sabour12}
\bibinfo{author}{\bibfnamefont{M.} \bibnamefont{Jafarpour}} \bibnamefont{and}
\bibinfo{author}{\bibfnamefont{A.} \bibnamefont{Sabour}},
\bibinfo{author}{\bibfnamefont{A useful strong lower bound on two-qubit concurrence}},
 \bibinfo{journal}{Quant. Inf. Process.} \textbf{\bibinfo{volume}{11}},
 \bibinfo{pages}{1389} (\bibinfo{year}{2012}).
 
\bibitem[{\citenamefont{Yu et al}(2007)}]{yu07}
\bibinfo{author}{\bibfnamefont{T.} \bibnamefont{Yu}} \bibnamefont{and}
\bibinfo{author}{\bibfnamefont{J. H.} \bibnamefont{Eberly}},
\bibinfo{author}{\bibfnamefont{Evolution from entanglement to decoherence of bipartite mixed X states}},
 \bibinfo{journal}{Quant. Inf. and Comp.} \textbf{\bibinfo{volume}{7}},
 \bibinfo{pages}{459} (\bibinfo{year}{2007}).  

\bibitem{comment2}
\bibinfo{journal}{Even though the inequality \eqref{E>S3}} may not hold for other measures of entanglement and steering, a hierarchy equivalent to $S_3 \Rightarrow E$ will still apply if such measures are monotonous functions of concurrence and $S_3$.

\bibitem[{\citenamefont{Collins et al}(2004)}]{collins04}
\bibinfo{author}{\bibfnamefont{D.} \bibnamefont{Collins}} \bibnamefont{and}
\bibinfo{author}{\bibfnamefont{N.} \bibnamefont{Gisin}},
\bibinfo{author}{\bibfnamefont{A relevant two qubit Bell inequality inequivalent to the CHSH inequality}},
 \bibinfo{journal}{J. Phys. A: Math. Gen.} \textbf{\bibinfo{volume}{37}},
 \bibinfo{pages}{1775} (\bibinfo{year}{2004}).
 
\bibitem[{\citenamefont{M\'ethot et al}(2007)}]{methot07}
\bibinfo{author}{\bibfnamefont{A. A.} \bibnamefont{M\'ethot}} \bibnamefont{and}
\bibinfo{author}{\bibfnamefont{V.} \bibnamefont{Scarani}},
\bibinfo{author}{\bibfnamefont{An anomaly of non-locality}},
 \bibinfo{journal}{Quantum Inf. Comput.} \textbf{\bibinfo{volume}{7}},
 \bibinfo{pages}{157} (\bibinfo{year}{2007}).
 

\end{thebibliography}
\end{document}